\documentclass[aps,pre,twocolumn,showpacs,floatfix,superscriptaddress]{revtex4}
\usepackage{graphicx}
\usepackage{amssymb}
\usepackage{amsmath}
\usepackage{lmodern}
\usepackage{color}
\usepackage{hyperref}
\usepackage{simplewick}
\begin{document}

\preprint{This line only printed with preprint option}

\title{Lift force due to odd (Hall) viscosity}

\author{E. Kogan}

\affiliation{Jack and Pearl Resnick Institute, Department of Physics, Bar-Ilan University, Ramat-Gan 52900, Israel}
\affiliation{Max-Planck-Institut fur Physik komplexer Systeme,  Dresden 01187, Germany}
\affiliation{Center for Theoretical Physics of Complex Systems, Institute for Basic Science (IBS), Daejeon 34051, Republic of Korea}

\date{\today}

\begin{abstract}
We study the problem of flow of a neutral gas past an infinite cylinder at right angle to its axis at low Reynolds number when the fluid is characterized
by broken time-reversal invariance, and hence by odd viscosity in addition to the normal even one. We solve the Oseen approximation to
  Navier-Stokes equation and calculate the lift force which appears due to the odd viscosity.
\end{abstract}


\maketitle

Viscosity, i.e. the resistance to a flow in which adjacent parts of a fluid move with different
velocities, is a basic property of all classical and quantum liquids, and becomes relevant for
electron liquids as well, when disorder and coupling to the lattice are not too strong.

In rotationally-invariant systems, when the time-reversal symmetry is not broken, the
viscosity tensor is entirely described by two scalar transport coefficients, the shear and the
bulk viscosities (denoted by $\eta$ and $\zeta$ respectively), which are both dissipative.

The Hall viscosity, also known as the Hall viscosity  and Lorentz shear modulus, is an off-diagonal viscosity
term that is dissipationless and produces forces perpendicular
to the direction of the fluid flow. It can have a quantum mechanical
origin in, for example, systems exhibiting the quantum
Hall effect \cite{avron0,avron,levay,read,hughes,read2,bradlyn,hoyos,hughes2,tokatly2,tokatly1,sherafati} or a classical origin in plasmas at
finite-temperature \cite{pitaevskii}. More generally, Hall viscosity presents  a link between quantum Hall systems, plasmas and liquid crystals \cite{lingam}.
We will not focus on the microscopic origin of the Hall viscosity
coefficient, but only assume it to be non-vanishing in
conjunction with the usual viscosity coefficients.

Despite the extensive theoretical discussions on its properties, the
question of how to measure the odd viscosity still poses a challenge. We will show that odd viscosity of a neutral gas can be determined from the measurement of
the lift force which acts due to flow at right angle to the axis of
infinite cylinder

Euler's equation can be written in the form
\begin{eqnarray}
\label{ll}
\frac{\partial}{\partial t}(\rho v_i)=-\frac{\partial\Pi_{ik}}{\partial x_k},
\end{eqnarray}
where $\rho$ is the fluid density, $v_i$ is the fluid velocity, and  $\Pi_{ik}$ is the momentum flux density tensor.
The equation of motion of a viscous fluid may be obtained by subtracting from the "ideal" momentum flux a viscous stress tensor $\sigma'_{ik}$
which gives the viscous transfer of momentum in the fluid \cite{landau}
\begin{eqnarray}
\Pi_{ik}=p\delta_{ik}+\rho v_iv_k-\sigma'_{ik}
\end{eqnarray}.

Considering the system with the symmetry, equivalent to those which characterizes isotropic system with magnetic field along the $z$-axis,
we can write down the components of the stress tensor as \cite{lifshitz}
\begin{eqnarray}
\label{pi}
\sigma'_{xx}&=&-\eta_0\left(V_{zz}-\frac{1}{3}\text{div}\;{\bf V}\right)+\eta_1\left(V_{xx}-V_{yy}\right)\nonumber\\
&+&2\eta_3V_{xy}+\zeta\; \text{div}\;{\bf V}+\zeta_1V_{zz}\nonumber\\
\sigma'_{yy}&=&-\eta_0\left(V_{zz}-\frac{1}{3}\text{div}\;{\bf V}\right)+\eta_1\left(V_{xx}-V_{yy}\right)\nonumber\\
&-&2\eta_3V_{xy}+\zeta \;\text{div}\;{\bf V}+\zeta_1V_{zz}\nonumber\\
\sigma'_{zz}&=&2\eta_0\left(V_{zz}-\frac{1}{3}\text{div}\;{\bf V}\right)+\zeta \;\text{div}\;{\bf V}+\zeta_1(V_{zz}+\text{div}\;{\bf V})\nonumber\\
\sigma'_{xy}&=&2\eta_1V_{xy}-\eta_3\left(V_{xx}-V_{yy}\right)\nonumber\\
\sigma'_{xz}&=&2\eta_2V_{xz}+2\eta_4V_{yz}\nonumber\\
\sigma'_{yz}&=&2\eta_2V_{yz}-2\eta_4V_{xz},
\end{eqnarray}
where
\begin{eqnarray}
V_{ij}=\frac{1}{2}\left(\frac{\partial v_i}{\partial x_j}+\frac{\partial v_j}{\partial x_i}\right).
\end{eqnarray}
Here the terms with the coefficients $\eta_3$ and $\eta_4$ describe odd viscosity.

As a side note, we would like to mention that Eq. (\ref{pi}) would be also valid for magnitoactive plasma.
 In this case the equations are even somewhat simplified by the vanishing of two second viscosity coefficients $\zeta$ and $\zeta_1$  \cite{lifshitz}.
For the reader to get a feeling of the physics behind the  viscosity coefficients introduced above, we would like to reproduce the results
obtained in the framework of the fluid dynamics in a magnitoactive plasma. In this case in the strong magnetic field one gets  \cite{lifshitz}
\begin{eqnarray}
\eta_1&=&\frac{\eta_2}{4}=\frac{2\pi^{1/2}(ze)^4L_iN_i^2}{5(MT)^{1/2}\omega_{Bi}^2} \\
\eta_3&=&\frac{\eta_4}{2}=\frac{N_iT}{2\omega_{Bi}},
\end{eqnarray}
where $T$ is the temperature, $M$ is the ion mass, $ze$ is the ion charge, $N_i$ is the ion concentration, $\omega_{Bi}=zeB/Mc$, $B$ is the magnetic field,
and $L_i$ is the Coulomb logarithm. However, for the case of plasma, if we assume, as it is normally done, that magnetic field is the source of the odd viscosity, we should include Lorentz force into Eq. (\ref{ll}). The results for this case will be reported separately.

Eq. (\ref{pi}) is simplified if we consider  two-dimensional (in the $xy$ plane) fluid motion, and additionally assume the fluid  incompressible (${\bf\nabla\cdot v} = 0$). Combining this equation with Eq. (\ref{ll}) we obtain
modified
 Navier-Stokes equation, which can be written in vector form \cite{avron}
\begin{eqnarray}
\label{ns}
\rho\left[\partial_t{\bf v} + ({\bf v\cdot\nabla}){\bf v}\right] =-{\bf \nabla}p + \eta_1\Delta{\bf v}+\eta_3\Delta{\bf v}^*,
\end{eqnarray}
where the vector ${\bf v}$ is two-dimensional, and the dual is defined, as usual, by
\begin{eqnarray}
\label{dual}
v_i^*=\epsilon_{ij}v_j.
\end{eqnarray}

Here probably it is appropriate to mention another aspect of Hall viscosity $\eta_3$ relevant for modern studies of the quantum Hall systems.
The Hall viscosity is an instance
of a class of "anomalous transport coefficients" - of which
the Hall conductivity is the best known example - which
are given by the imaginary part of an off-diagonal linear
response function \cite{sherafati}, in this case \cite{tokatly2}
\begin{eqnarray}
\eta_3=\lim_{\omega\to 0}\text{Im}\frac{<<P_{xx};P_{xy}>>_{\omega}}{\omega},
\end{eqnarray}
where $<<P_{xx};P_{xy}>>_{\omega}$ is a shorthand for the off-diagonal
stress-stress response function.

Returning to Eq. (\ref{ns}) we realize that equation ${\bf\nabla\cdot v} = 0$ allows to introduce the stream function $\psi$
\begin{eqnarray}
\label{ns45}
v_x=\frac{\partial \psi}{\partial y},\;\;\;v_y=-\frac{\partial \psi}{\partial x}.
\end{eqnarray}
Noticing that
\begin{eqnarray}
v^*=-\text{grad}\;\psi,
\end{eqnarray}
we can rewrite Eq. (\ref{ns}) in the form \cite{avron}
\begin{eqnarray}
\label{ns0}
\rho\left[\partial_t{\bf v} + ({\bf v\cdot\nabla}){\bf v}\right] =-{\bf \nabla}(p+\eta_3\Delta\psi) + \eta_1\Delta{\bf v}.
\end{eqnarray}

The pressure can be eliminated from Eq. (\ref{ns0})  by taking the curl of both sides. We get
the well known equation
\begin{eqnarray}
\label{ns4}
\partial_t(\text{curl}\;{\bf v}) + ({\bf v\cdot\nabla})(\text{curl}\;{\bf v})-(\text{curl\;}{\bf v}\cdot {\bf \nabla}){\bf v}\nonumber\\
 = \nu\Delta(\text{curl}\;{\bf v}).
\end{eqnarray}
When the velocity distribution is known, the pressure distribution in the fluid can be found from Eq. (\ref{ns0}).
Substituting $\psi$ for ${\bf v}$ we get
 \cite{landau}
\begin{eqnarray}
\label{ns5}
\frac{\partial}{\partial t}\Delta \psi-\frac{\partial \psi}{\partial x}\frac{\partial \Delta\psi}{\partial y}+
\frac{\partial \psi}{\partial y}\frac{\partial \Delta\psi}{\partial x}-\nu\Delta\Delta\psi=0.
\end{eqnarray}

 It is evident from Eq. (\ref{ns0}) that typically odd viscosity
does not influence the distribution on velocity, but does influence distribution of pressure.
Additional pressure appearing in the system due to odd viscosity can be presented as
\begin{eqnarray}
\Delta p=\eta_3\left(\text{curl}\;{\bf v}\right)_z.
\end{eqnarray}
Thus odd viscosity, at least in principle, can be measure by measuring forces the fluid apply to the bodies it flows past.

To warm up,
consider two elementary problems.
Consider first,
steady flow between two fixed parallel lines in the presence of a pressure gradient.
We choose the $x$-axis in the direction of the motion of the fluid, and $y$-axis in the perpendicular direction,
so the lines are given by equations $y=0$ and $y=h$.
Since the velocity clearly is in the $x$ direction and does not depend upon $x$,
we can look for $\psi$ in the form
$\psi=\psi(y)$, and
Eq. (\ref{ns5}) gives
\begin{eqnarray}
\label{ns6}
\frac{d^4 \psi}{dy^4}=0.
\end{eqnarray}
The two boundary conditions are
\begin{eqnarray}
\label{ns7}
\left.\frac{d\psi}{dy}\right|_{y=0}=\left.\frac{d\psi}{dy}\right|_{y=h}=0.
\end{eqnarray}
The solution of Eq. (\ref{ns6}) is
\begin{eqnarray}
\psi=Ay^2\left(\frac{3}{2}h-y\right)
\end{eqnarray}
(the constant term in $\psi$ is irrelevant). Thus for velocity we obtain
\begin{eqnarray}
v_x=3Ay(h-y)
\end{eqnarray}
and
\begin{eqnarray}
p=-6A\left(\eta_1 x-{\eta_3}y\right)+\text{const}.
\end{eqnarray}

To consider our next elementary example, motion of a fluid between two coaxial circles rotating with radii $R_1$, $R_2$ ($R_2>R_1$), rotating about their axis
with angular velocities $\Omega_1$, $\Omega_2$,
we need to rewrite Eqs. (\ref{dual}), (\ref{ns45})  and (\ref{ns5}) in polar coordinates. The dual becomes
\begin{eqnarray}
v^*_{\theta}=-v_r,\;\;\;v_r=v_{\theta},
\end{eqnarray}
Eq. (\ref{ns45}) becomes
\begin{eqnarray}
\label{ns46}
v_\theta=\frac{\partial \psi}{\partial r},\;\;\;v_r=-\frac{1}{r}\frac{\partial \psi}{\partial \theta}.
\end{eqnarray}
and Eq. (\ref{ns5}) becomes
\begin{eqnarray}
\label{ns10}
\frac{\partial}{\partial t}\Delta \psi-\frac{1}{r}\left(\frac{\partial \psi}{\partial r}\frac{\partial \Delta\psi}{\partial \theta}+
\frac{\partial \psi}{\partial \theta}\frac{\partial \Delta\psi}{\partial r}\right)-\nu\Delta\Delta\psi=0.
\end{eqnarray}
From the symmetry we have $\psi=\psi(r)$, and Eq. (\ref{ns10}) gives
\begin{eqnarray}
\label{ns9}
\frac{1}{r}\frac{d}{dr}\left\{r\frac{d}{dr}\left[\frac{1}{r}\frac{d}{dr}\left(r\frac{d\psi}{dr}\right)\right]\right\}=0
\end{eqnarray}
The two boundary conditions are
\begin{eqnarray}
\left.\frac{d\psi}{dr}\right|_{r=R_1}=\Omega_1R_1^2,\;\;\;\left.\frac{d\psi}{dr}\right|_{r=R_2}=\Omega_2R_1^2.
\end{eqnarray}
The solution of Eq. (\ref{ns9}) is
\begin{eqnarray}
\psi=\frac{\Omega_2R_2^2-\Omega_1R_1^2}{R_2^2-R_1^2}\frac{r^2}{2}+\frac{(\Omega_1-\Omega_2)R_1^2R_2^2}{R_2^2-R_1^2}\ln r.
\end{eqnarray}
Hence, for velocity we obtain  \cite{landau}
\begin{eqnarray}
v_{\phi}=\frac{\Omega_2R_2^2-\Omega_1R_1^2}{R_2^2-R_1^2}r+\frac{(\Omega_1-\Omega_2)R_1^2R_2^2}{R_2^2-R_1^2}\frac{1}{r}.
\end{eqnarray}
The pressure is
\begin{eqnarray}
p=\rho\frac{v_{\phi}^2}{r}-\eta_3\Delta\psi+\text{const}.
\end{eqnarray}
Thus we see that odd viscosity does not influence ether velocity distribution or the applied torque,
 but just
pressure on the circles, which are proportional to the rate of rotation (this fact can be seen directly from Eq. (\ref{ns0})).

Now consider the lift force on the cylinder of radius $a$ moving in a fluid with velocity ${\bf U}$, corresponding to low Reynolds number.
(Because the problem is essentially two-dimensional we'll use Eq. (\ref{ns}).)
 The natural desire would be to ignore
the term $({\bf v\cdot\nabla}){\bf v}$ in this equation. However, it is well known that (in the absence of odd viscosity) for the problem considered,
such amputated equation does not have a solution \cite{lamb,landau}. The way out was discovered by Oseen and Lamb \cite{lamb}.
It consists in approximating  Navier-Stokes equation
\begin{eqnarray}
\rho\left[\partial_t{\bf v} + ({\bf v\cdot\nabla}){\bf v}\right] =-{\bf \nabla}p + \eta_1\Delta{\bf v}
\end{eqnarray}
 by the equation
\begin{eqnarray}
\label{oseen}
\rho ({\bf U\cdot\nabla}){\bf v} =-{\bf \nabla}p + \eta_1\Delta{\bf v}.
\end{eqnarray}
Hence, in our case one has to solve equation
\begin{eqnarray}
\label{oseen2}
\rho ({\bf U\cdot\nabla}){\bf v} =-{\bf \nabla}p + \eta_1\Delta{\bf v}+\eta_3\Delta{\bf v}^*,
\end{eqnarray}
or alternatively
\begin{eqnarray}
\label{oseen3}
\rho ({\bf U\cdot\nabla}){\bf v} =-{\bf \nabla}(p+\eta_3\Delta\psi) + \eta_1\Delta{\bf v}.
\end{eqnarray}

Let us recall classical results due to Lamb (in the absence of odd viscosity) \cite{lamb}.
Going to the frame of reference moving with the circle, one obtains that Eq. (\ref{oseen}) is solved by the substitutions
($x$ axis is chosen in the direction of ${\bf U}$):
\begin{eqnarray}
v_x&=&-\frac{\partial \phi}{\partial x}+\frac{1}{2k}\frac{\partial \chi}{\partial x}-\chi\\
v_y&=&-\frac{\partial \phi}{\partial y}+\frac{1}{2k}\frac{\partial \chi}{\partial y}\\
p&=&\rho U \frac{\partial \phi}{\partial x}
\end{eqnarray}
where $k=U/2\nu$, provided that functions $\phi$ and $\chi$ satisfy equations
\begin{eqnarray}
\Delta\phi&=&0\\
\label{4}
\left(\Delta-2k\frac{\partial}{\partial x}\right)\chi&=&0.
\end{eqnarray}
The solution of Eq. (\ref{4}) is
\begin{eqnarray}
\chi=Ce^{kx}K_0(kr).
\end{eqnarray}
where $K_0$ is the modified Bessel function. For small values of $kr$ we have
\begin{eqnarray}
\chi=-C(1+kx)\left(\gamma + \ln\left(\frac{1}{2}ka\right)\right),
\end{eqnarray}
where $\gamma$ is  Euler' constant. Hence for these values
\begin{eqnarray}
&&\frac{1}{2k}\frac{\partial \chi}{\partial x}-\chi=-\frac{C}{2k}\left\{k\left[\frac{1}{2}-\gamma-\ln\left(\frac{1}{2}ka\right)\right]\right.\nonumber\\
&&+\left.\frac{\partial}{\partial x}\ln r-\frac{1}{2}kr^2\frac{\partial^2}{\partial x^2}\ln r+\dots\right\} \\
&&\frac{1}{2k}\frac{\partial \chi}{\partial y}=-\frac{C}{2k}\left\{
\frac{\partial}{\partial y}\ln r-\frac{1}{2}kr^2\frac{\partial^2}{\partial x\partial y}\ln r+\dots\right\}.\nonumber\\
\end{eqnarray}
If we set
\begin{eqnarray}
\phi=A_0\ln r +A_1\frac{\partial}{\partial x}\ln r+\dots,
\end{eqnarray}
than we'll find that conditions $v_x=-U$, $v_y=0$ at $r=a$ will be satisfied if we set approximately
\begin{eqnarray}
C&=&\frac{2U}{\frac{1}{2}-\gamma-\ln\left(\frac{1}{2}ka\right)}\\
A_0&=&-\frac{C}{2k}\\
A_1&=&\frac{1}{4}Ca^2
\end{eqnarray}
The vorticity is thus given by the equation \cite{lamb}
\begin{eqnarray}
\left(\text{curl}\;{\bf v}\right)_z=-kC\frac{y}{r}e^{kx}K_1(kr),
\end{eqnarray}
Taking into account that $ka\ll 1$ we obtain at the surface of the cylinder
\begin{eqnarray}
\left(\text{curl}\;{\bf v}\right)_z=-\frac{C}{a}\sin\theta.
\end{eqnarray}
Thus in the approximation considered
\begin{eqnarray}
\Delta F_x&=&0\\
\Delta F_y&=&-\pi\eta_3C.
\end{eqnarray}
Notice, that
the  drag force on the unit of length of the cylinder \cite{lamb}
\begin{eqnarray}
F_x=2\pi\eta_1 C.
\end{eqnarray}

To conclude we would like to emphasize again that the calculated lift force appears exclusively due to the non-dissipative odd viscosity, and hence can exist
 only if the time reversal invariance in the system is broken, either explicitly or implicitly.
 
 We  also want to mention that the obtained  results can be applied, in addition to neutral gas, mentioned in the beginning of the paper,
 to other systems, like, for example, nematic liquid crystals.

\begin{acknowledgments}

Discussions with M. Sherafati, which actually   initiated the present work, are gratefully acknowledged.

The authors also cordially thanks  for the hospitality extended to him during
his stay: Max-Planck-Institut fur Physik komplexer Systeme, where the work was initiated, and
Center for Theoretical Physics of Complex Systems, where the work continued.

\end{acknowledgments}

\end{document}